\documentclass[12pt]{iopart}
\usepackage{times}
\usepackage{graphicx}
\usepackage{epsfig}

\newcommand{\AmS}{{\protect\the\textfont2
  A\kern-.1667em\lower.5ex\hbox{M}\kern-.125emS}}

\newcommand{\ber}{\begin{eqnarray}}
\newcommand{\eer}{\end{eqnarray}}
\newcommand{\be}{\begin{eqnarray}}
\newcommand{\ee}{\end{eqnarray}}

\def\Toprel#1\over#2{\mathrel{\mathop{#2}\limits^{#1}}}

\begin{document}

\title{Group projection method in statistical systems }
\author{Krzysztof Redlich$^{~a,b}$, Frithjof Karsch$^{~a}$ and  Ahmed Tounsi$^{~c}$ %and Ludwik
%Turko$^{b}$}
}
\address{$^a$ Fakult\"at f\"ur Physik, Universit\"at
Bielefeld, Postfach 100 131, D-33501 Bielefeld, Germany\\ $^b$
Institute of Theoretical Physics, University of Wroc\l aw,
PL-50204 Wroc\l aw,
Poland \\
%\author{}
%\address{ Fakult\"at f\"ur Physik, Universit\"at Bielefeld,
%Postfach 100 131, D-33501 Bielefeld, Germany}
%$\author{Ahmed Tounsi} $\address
 $^c$ Laboratoire de Physique
Th\'eorique et Hautes Energies, Universit\'e Paris 7, Paris,
France}
%\author{Ludwik Turko}
%\address{ Institute of
%Theoretical Physics, University of Wroc\l aw, PL-50204 Wroc\l aw,
%Poland }

\begin{abstract}
We discuss an application of group theoretical methods to the
formulation of the thermodynamics  of systems constrained by the
conservation laws described by a semi--simple compact Lie group. A
general projection method that allows to construct a partition
function for a given  irreducible representation of the Lie group
is outlined. Applications of the method in  Lattice Gauge Theory
(LGT) for non--zero baryon number and in  the phenomenological
description of particle production in ultrarelativistic heavy ion
collisions are also indicated.

\end{abstract}
\section{Introduction}

In the formulation of the  thermodynamics of a strongly
interacting medium  one needs in general to implement the
constraints imposed by the conservation laws that are governed  by
an internal symmetry of the Hamiltonian
\cite{r38,r381,r64,r64n,rgen,peter,tounsi,r20}. In this paper, we
discuss how applying  the basic properties of the Lie groups and
their representations one can derive the partition function that
accounts for these constraints. We present examples of the
application of the method for different statistical systems and
phenomenological models.

\section{Projected partition function}

The usual way of treating the problem of quantum number
conservation in  statistical physics is by introducing the
grand canonical partition function,

\be
Z(\mu_S,T,V)={\rm Tr}[e^{-\beta (\hat{H}-\mu_{S}  \hat{ S})}],
\label{eq40} \ee
with $\hat H$ being the Hamiltonian, $\hat S$ the charge
operator, $\beta$  the inverse temperature, $V$ the volume of
the system  and $\mu_S$ the chemical potential associated with the
conserved charge  $S$. The chemical potential in Eq.~(\ref{eq40})
plays the role of the Lagrange multiplier which is
% that guarantees charge conservation.
%The
%$\mu_S$
 fixed by the condition that the charge
  is conserved on the average and has the required
value $\langle S\rangle$ such that:
\be \langle S\rangle=T { {\partial\ln Z(\mu_S,T)}\over {\partial
\mu_S}} \label{eq41}. \ee

The grand canonical partition  function (\ref{eq40}) provides an
adequate description of the statistical properties of the system
only if the number of particles carrying charge $S$ is
asymptotically  large and if its fluctuations can be neglected.

In order to derive a more general  statistical operator that is
free from the above requirements one  usually replaces the
function (\ref{eq40}) by the  canonical partition function $Z_S$
that accounts for an exact charge conservation
 \be
 Z_S(T,V):={\rm Tr_S}[e^{-\beta \hat H}].
 \label{eq43} \ee
The subscript   $S$ under the trace indicates that it is
restricted to
 the states that carry   an exact
   value $S$ of the conserved charge. Obviously $Z_S(T,V)$ from
  Eq.~(\ref{eq43}) and $Z(\mu_S,T,V)$  from Eq.~(\ref{eq40}) are connected via
  cluster decomposition

 \be Z(\mu_S,T,V)=\sum_{s=-\infty}^{s=+\infty} Z_S \lambda_S^s,
 \label{eq42} \ee
with the fugacity parameter $\lambda_S:=e^{\beta\mu_S}$.
 Thus, $Z_S$ is viewed  as a coefficient in the Laurent series of $Z$ in the
fugacity. Applying the Cauchy formula in Eq.~(\ref{eq42}) one
calculates  $Z_S$ by taking an inverse transformation:

%\be Z_S(T,V)={{1}\over {2\pi i}}\oint {{d\lambda_S}\over
%{\lambda_S^{s+1}}} Z(\lambda_S,T,V) \label{eq44} \ee
%
%or choosing the integration path as the unit circle
%$\lambda_S=\exp{(i\phi)}$ one converts the  contour  into the
%angular integral
%
\be Z_S(T,V)=\int_{-\pi}^{+\pi} {{d\phi}\over {2\pi}} e^{-is\phi}
\tilde Z(\phi ,T,V) \label{eq45}.
 \ee
The generating function $\tilde Z(\phi,T,V):= Z(\lambda_S
=e^{i\phi },T,V)$
is  obtained
from the grand canonical partition function (\ref{eq40}) by  "Wick
rotation" of the chemical potential, $\mu_S\to i\phi$. The function
$\tilde Z$ is unique  for all canonical partition functions that
are formulated for a  fixed value of the conserved charge.

The integral (\ref{eq45}) describes  the projection   onto
the canonical partition function that accounts for an exact
conservation of an abelian charge. This is the projection
procedure as $Z_S$ is also obtained from
\be
Z_S(T,V)=
%{\rm Tr_S}[e^{-\beta \hat H}]
  {\rm Tr}[e^{-\beta \hat
 H}\hat P_S] \label{eq46} \ee
  where $\hat P_S$ is the projection operator on
the states with a given value  of the conserved  charge  $S$. For
an additive quantum numbers, $\hat P_S$ is just  a delta function
$P_S=\delta_{ S^,,S}\hat I$. Using the Fourier expression of
$\delta_{ S^,,S}$  in Eq.~(\ref{eq46}) one  reproduces the
projected formula (\ref{eq45}).

The conservation of an additive quantum number
 is usually related with an
invariance of the Hamiltonian under the abelian  U(1) internal
symmetry group. In many physics  applications it is of importance
to generalize the projection method to symmetries that are related
with a non-abelian Lie groups $G$. An example is a special unitary
group SU(N) that plays an essential role in the strong
interaction. A generalization of the projection method would
require to specify the projection operator or the generating
function. Consequently, the partition function obtained with a
specific eigenvalue of the Casimir operator that fixes  the
multiplet of the irreducible representation of the symmetry group
$G$ can be determined.

To find the generating function for the  canonical partition
function with respect to the symmetry group  $G$, one  introduces
  $\tilde Z(g)$ \cite{r38}

 \be
 \tilde Z(g):={\rm Tr }[U(g)e^{-\beta\hat H}] \label{gen}
\label{eq47}
 \ee
as  the   function on the group $G$ with U(g) being a unitary
representation of the group and $g\in G$. The U(g) can be
decomposed into irreducible representations $U_\alpha(g)$,
 \be
 U(g)=\sum_\alpha^\bigoplus U_\alpha(g),
\label{eq48}
 \ee
and thus, from Eqs.~(\ref{eq47}) and (\ref{eq48}) one can write
explicitly
 \ber
 \tilde Z(g):&=&\sum_\alpha{\rm Tr_\alpha }[U_\alpha (g)e^{-\beta\hat H}]
 \nonumber  \\
&=&\sum_\alpha\sum_{\nu_\alpha,\xi_\alpha} \langle
\nu_\alpha,\xi_\alpha\mid U_\alpha (g)e^{-\beta\hat H}\mid
\nu_\alpha,\xi_\alpha\rangle,
 \label{gen2}
\label{eq49}
 \eer
where $\nu_\alpha$ labels the states  within the representation
$\alpha$ and  $\xi_\alpha$ are degeneracy parameters of a given
representation. Due to the requirement of an  exact  symmetry the
only non--vanishing matrix elements of the evolution operator
$e^{-\beta \hat H}$ are those diagonal in $\nu_\alpha$. The matrix
elements of $U_\alpha(g)$ are
 non-zero if they are diagonal in $\xi_\alpha$. Finally, the
matrix elements of the Hamiltonian are independent of the states
within representation and those of $U(g)$ of degeneracy factors.
Consequently, the matrix  elements in Eq.~(\ref{eq49}) factorize
and the generating function is

\be
 \tilde Z(g)=\sum_\alpha {{\chi_\alpha(g)}\over {d(\alpha)}}
Z_\alpha(T,V),
 %\label{las}
\label{eq53}
 \ee
where $d(\alpha)$ is the  dimension  and $\chi_\alpha$ is the
character of the irreducible representation  $\alpha$. In the
above expression  $Z_\alpha$ is  introduced as  a {\it canonical
partition function} with respect to $G$ symmetry and is defined as

 \be
%{ {1}\over {d(\alpha)}}
Z_\alpha(T,V):=%
 {\rm
Tr_\alpha} e^{-\beta\hat H}= {  {d(\alpha)}}\sum_{\xi_\alpha}
\langle \xi_\alpha\mid e^{-\beta\hat H}\mid \xi_\alpha\rangle.
 \label{gen6}
\label{eq52}
 \ee

%Calculating $Z_\alpha$ one restricts under the trace only those
%states that transform with respect to a given irreducible
%representation of the symmetry group.

  Eq.~(\ref{eq53}) connects a canonical partition function
$Z_\alpha$ to the generating function on the group. Thus,
$Z_\alpha$ is the coefficient in the cluster decomposition of the
generating function with respect to the characters of the
representations associated with the symmetry group.

\noindent
%The character functions satisfy the orthogonality relation
%%
%\be
%  {1\over {d(\alpha})}\int
%  d\mu(g)\chi_\alpha^*(g)\chi_\gamma(g)=\delta_{\alpha,\gamma}
%  \label{or}
%\label{eq54}
% \ee
%
%where $d\mu(g)$ is an invariant Haar measure on the group.
The orthogonality relation of the  characters,
% ${1/
%{d(\alpha})}\int
% d\mu(g)\chi_\alpha^*(g)\chi_\gamma(g)=\delta_{\alpha,\gamma}$
\be
  {1\over {d(\alpha})}\int
  d\mu(g)\chi_\alpha^*(g)\chi_\gamma(g)=\delta_{\alpha,\gamma}
  \label{or}
\label{eq54}
 \ee
allows to find  the canonical partition function.
 From (\ref{eq53}) and (\ref{eq54})
one gets
 \be
Z_\alpha(T,V)=d(\alpha)\int d\mu(g) {\chi_\alpha^*(g)} \tilde
Z(g).
  \label{maing}
\label{eq55}
 \ee
This result is a generalization of (\ref{eq45}) to an arbitrary
internal symmetry group that is a compact Lie group. The formula
holds  for any dynamical system as it is independent of the
specific form of the   Hamiltonian.

 To find the
canonical partition function one needs  to determine  first the
generating function $\tilde Z(g)$ defined on   the symmetry group
$G$. If the
 group is of rank $r$, then the character of any irreducible
representation   is a function of $r$ variables
$\{\gamma_1,_{....},\gamma_r\}$, thus also is the generating
function (10).
%
%(\ref{eq55}).
Diagonalizing the unitary operators $U(g)$ under the trace in
Eq.(\ref{eq47}) and denoting by $J_k$ ($k=1,...,r$) the commuting
generators of $G$,
%and
%the with the character function
%
% \be
%
%\chi_\alpha(\gamma_1,..,\gamma_r)=\sum_{\nu_\alpha} \langle
%\nu_\alpha\mid e^{i\sum_{i=1}^{i=r}\gamma_iJ_i}\mid \nu_\alpha
%\rangle
%  \label{las1}
%\label{eq56}
% \ee
%
%where $\nu_\alpha$ labels  the state within the representation
%$\alpha$. With the above  form of the characters we can write
%Eq.~(\ref{eq53}) as
%
the generating function can be  formulated on the maximal abelian
subgroup of G as
 \be
\tilde Z(\gamma_1,..,\gamma_r)={\rm Tr}[ e^{-\beta\hat H+
i\sum_{i=1}^{r}\gamma_iJ_i}].
  \label{la1}
\label{eq57}
 \ee
Thus $\tilde Z(\vec \gamma)$
%
%Through the Wick rotation $\gamma_i=-i\beta\mu_i$
%the generating function $\tilde Z$
is just the GC partition function with  complex chemical
potentials that are associated with all generators of   the Cartan
sub-algebra.

  Equations (\ref{eq55}) and (\ref{eq57}) provide  a complete
description of the statistical operators that account for the
constraints imposed by the conservation laws of internal
symmetries of the Hamiltonian.
%the basis that permits to obtain the canonical partition function
%for systems restricted to any symmetry.
The simplicity of the projection formula (\ref{eq55}) is   that
the operators that appear in the generating function are additive.
One sees that the problem of extracting the canonical  partition
function with respect to an arbitrary compact Lie group $G$ is
reduced to the projection onto a maximal abelian subgroup of $G$.

The  generating function (\ref{eq57}) can be calculated by
applying  standard perturbative diagrammatic methods or a mean
field approach.  However,  if the interactions in the Hamiltonian
can be omitted or can effectively be  described as a
modification of the particle dispersion relations by implementing
an effective particle mass, then the trace in Eq.~(\ref{eq57}) can
be done exactly. In this particular situation the generating
function can be written \cite{r38} as

 \be
\tilde Z(\vec \gamma)=\exp [{\sum_\alpha {{\chi_\alpha(\vec
\gamma)}\over {d(\alpha)}}Z^1_\alpha}]
%  \label{la3}
\label{eq58} \ee
%
%where $\vec \gamma =(\gamma_1,_..,\gamma_r)$ and
where the one-particle partition function in Boltzmann approximation
 \be
 Z^1_\alpha=V\int {{d^3p}\over {(2\pi)^3}}
 \exp{(-\beta\sqrt{p^2+m_\alpha^2})} %n_{\alpha}(p)
 \ee
is just a thermal phase--space that is  available to all particles
of mass $m_\alpha$ that  belong  to a given irreducible multiplet.
% and $n_\alpha$ is the
%Boltzmann momentum distribution function.
The sum is taken over all particle representations that are 
constituents of a thermodynamical system.

\subsection{Phenomenological model of colour confinement}
\par
\noindent To illustrate how the projection method results in the
partition function, we discuss a statistical model that accounts
for the  conservation of a non-abelian charge related with the
global $G=SU_c$(N)$\times U_B(1)$ internal symmetry of the
Hamiltonian. As a physical system one considers  a thermal
fireball that is composed of quarks and gluons that carry the
colour degrees of freedom related with  $SU_c(N)$ symmetry
\cite{r64,r64n} and baryon number related with  $U_B(1)$ subgroup.
The system has temperature $T$ and volume $V$. The interactions
between quarks and gluons  are implemented effectively as
resulting in dynamical particle masses that are $T$ dependent,
e.g. through $m_{q,(g)}\sim g T$.   Thus, since under these
conditions the  free--particle dispersion relations are preserved,
Eq.~(\ref{eq58}) provides a correct description of the generating
function on the symmetry group G. The sum in the exponent in
(\ref{eq58}) gets the contributions from quarks, anti--quarks  and
gluons that transform respectively under the fundamental (0,1),
its conjugate (1,0) and adjoint (1,1) representation of the
$SU_c(N)$ symmetry group. Thus \cite{r64},
 \be
\ln \tilde Z(T,V,\vec\gamma,\gamma_B) = {{\chi_Q}\over {d_Q}}Z_Q^1
 + {{\chi_Q^*}\over {d_Q}}Z_{\bar Q}^1 + {{\chi_G}\over
{d_G}}Z_G^1
  \label{las4}
\label{eq59}
 \ee
where $\vec\gamma =(\gamma_1,..,\gamma_{N-1})$ are the parameters
of the  $SU_c(N)$ and $\gamma_B$ of the   $U_B(1)$ symmetry groups.
Through an explicit calculation of  one-particle partition
functions for massive quarks  the corresponding
generating functions  read:
 \ber
\ln\tilde Z_Q(T,V,\vec\gamma,\gamma_B)&=&\frac {g_Q}{d_Q}\frac
{m_Q^2VT}{2\pi^2}\sum_{n=0}^\infty  \frac {(-1)^{n+1}}{{n^2}}
K_2(nm_Q/T)
\nonumber\\
&[&e^{i\gamma_Bn/T}\chi_Q(n\vec \gamma)+
e^{-i\gamma_Bn/T}\chi_Q^*(n\vec \gamma) ]\label{las6} \label{eq60}
 \eer
where the two  terms in the brackets represent the contribution of
quarks and antiquarks respectively. The corresponding contribution for
massive gluons is obtained as
 \ber
\ln\tilde Z_G(T,V,\vec\gamma,\gamma_B)=\frac {g_G}{d_G}\frac
{m_G^2VT}{2\pi^2}\sum_{n=0}^\infty         \frac {1}{{n^2}}
K_2(nm_G/T)
%\nonumber\\
[\chi_G(n\vec \gamma)+ \chi_Q^*(n\vec \gamma) ]~~~~\label{las7}
\label{eq61}
 \eer
The coefficients $g_G,g_Q$ and $d_Q=N,d_G=N^2-1$, are respectively, the quark
and gluon spin--isospin degeneracy factors and dimensions of the
representations.

%\vspace{-4mm}
\begin{figure}[htb]
\begin{tabular}{cc}
\begin{minipage}{.6\textwidth}
%exe plot#jpsi-npart-sps
\centering\includegraphics[width=0.9
\textwidth,height=.75\textwidth,angle=180]{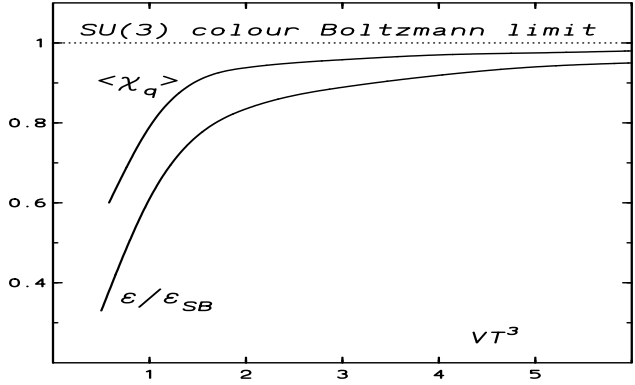}
\end{minipage}
&\hskip -2.0cm \begin{minipage}{.47\textwidth} \vspace{-5mm}
\caption{\label{fig1}    Thermodynamical functions obtained from
the projected partition function (\ref{eq64})  are compared to
their Stefan--Boltmann limit as a function of $VT^3$. The results
for the energy density $\epsilon/\epsilon_{SB}$ and the
expectation value of the character in the fundamental
representation of $SU(3)$ are indicated in the figure. The results
are obtained for a vanishing baryon--chemical potential $\mu_B$ in
(\ref{eq64}). }
%taking $N_{ch}=1530$ (NA49) leads to
\end{minipage}
\end{tabular}
\vskip -1cm
\end{figure}

The statistical operator is now obtained from (13) % (\ref{eq60})
through the projection of the generating functions (\ref{eq60})
and (\ref{eq61}) onto a given sector of quantum numbers. Here we
quote the results for the $SU_C(3)$ colour singlet partition
function that represents a global colour neutrality condition
(phenomenological confinement) of the quark--gluon plasma droplet
\cite{r64}:

 \ber
Z_{\alpha=0}(\mu_B,T,V)=\int d\mu(\gamma_1,\gamma_2) &&\exp
\{c_1\chi_G+ \nonumber \\ &&
c_2[L_R\cosh(\beta\mu_B)+iL_I\sinh(\beta\mu_B)] \}.
  \label{las10}
\label{eq64}
 \eer
where a  finite average value of the baryon number  is controlled
by the chemical potential $\mu_B$. The constants $c_1$ and $c_2$
can be extracted from Eqs.~(\ref{eq60}--\ref{eq61}) and  taking
only the first terms in the series.  The parameters  $L_I$, $L_R$
denotes the imaginary and real part of the character in the
fundamental representation of the $SU_C(3)$ group whereas $\chi_G$
is the character of the adjoint representation. The integration is
done on the $SU_C(3)$ group with an appropriate Haar measure $\mu
(\vec \gamma)$.

To illustrate how the group projection influence the
thermodynamics of the system we display  in Fig.~1 the behavior of
the  energy density and the thermal average of the characters as
obtained from the colour singlet statistical operator from
Eq.~(\ref{eq64}). The results are shown as a function of
dimensionless  parameter $VT^3$. In the absence of group
projection and for a vanishing value of $\mu_B$ the partition
function would have a simple form; $ Z(T,V)=\exp [c_1+c_2]$. The
resulting energy density $\epsilon_{SB}$ describes a well known
Stefan-Boltzmann limit and $\langle\chi_q\rangle =1$ in this case.
For an asymptotically large value of $VT^3\to \infty$, that
corresponds to the thermodynamical limit, the colour projected
results coincide with their Boltzmann values. However, for small
values of $VT^3$ a large suppression of thermodynamical quantities
is seen in Fig. 1. This is a generic feature of the partition
function restricted to a fixed representation. The requirement of
an exact conservation of the quantum numbers impose a strong
constraint on the particle thermal phase--space that results in
the suppression seen in Fig.~1. In the thermodynamical limit, the
number of particles that carry a quantum number related with a
given representation is so large that the  above constraints are
irrelevant.

%\vspace{-4mm}
\begin{figure}[htb]
\begin{tabular}{cc}
\begin{minipage}{.6\textwidth}
%exe plot#jpsi-npart-sps
\centering\includegraphics[width=1.\textwidth,height=.60\textwidth]{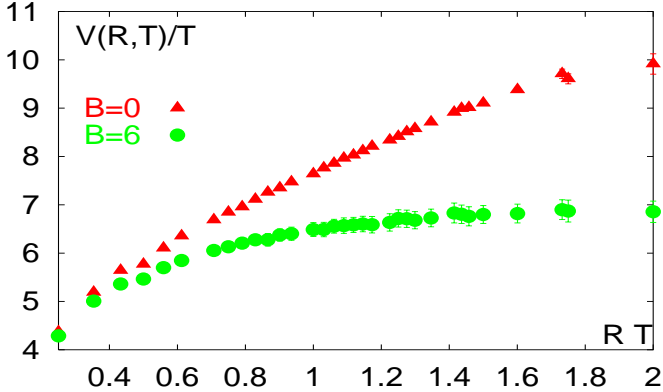}
\end{minipage}
&\hskip -2.0cm \begin{minipage}{.47\textwidth} \vspace{-11mm}
\caption{\label{fig2} Heavy quark potential from Ref. \cite{lgt4}
 as a function of quark anti--quark  separation $RT$. The results
were  obtained from Monte--Carlo simulations of quenched QCD at
fixed value of the coupling constant $\beta=5.62$ and  baryon
number $B=0$ and $B=6$.
 }
%taking $N_{ch}=1530$ (NA49) leads to
\end{minipage}
\end{tabular}
\vskip -0.5cm
\end{figure}

In  finite temperature gauge theory the zero component of the
gauge field $A_0$ takes on the role of the Lagrange multiplier
which guaranties  that all states satisfy the Gauss law
\cite{r64,lw,r64p}. In the Euclidean space one can choose a gauge
in such a way that $A_0^\nu(x,\tau)\lambda_\nu$ is a constant in
Euclidean time, so that
$
A_0^{ab}=g^{-1}\alpha^{ab}\delta_{ab}.
%  \label{las}
%\label{eq65}
$
In such a gauge the Wilson loop defined as
 \be
L(x)={1\over N} {\rm Tr}P\exp[ig\int_0^\beta A_0(x,\tau)d\tau]
\label{eq66}
 \ee
represents the character of the fundamental representation of the
$SU_c(N)$ group \cite{r64}. Thus,  the projected partition
function could be viewed  as an effective model that connects the
coloured quasi-particles \cite{quasi} with the Wilson loop
\cite{r64p}. This partition function can also be related to the
strong coupling effective free energy of the lattice gauge theory
with a finite chemical potential \cite{r64} and to the effective
potential of the $SU(N)$ spin model for the Wilson loop.

\subsection{ Projected partition function in  Lattice Gauge Theory}

The formulation of a finite baryon number density in QCD on the
lattice leads to a sever numerical problem of complex probability.
The QCD partition function formulated on the Euclidean lattice and
after  integrating out the Wilson fermions,  becomes
\cite{lgt,lgt1}
 \be
 Z(\mu_B,T,V)=\int \prod dU e^{S_G}\det[\aleph (\mu_B,U)]
 \ee
where $S_G$ is the gluon action, $\aleph:=1-\kappa M(U,\mu_B)$ is
the fermion matrix and $\kappa$ is the hopping parameter.  In
order to perform the  Monte-Carlo simulations with this
statistical operator it is necessary that the measure be real and
positive \cite{lgt1,lgt2,lgt3}. One situation where $\det[\aleph]$
is real is when there exist an invertible operator $P$ such that
$\aleph^+=P\aleph P^{-1}$. For the Wilson fermions at zero
chemical potential and for complex $\mu_B$  this relation hold
with $P=\gamma_5$. However, for real and non-vanishing $\mu_B\neq
0$ this is not anymore valid and the probabilistic interpretation
of the path integral representation of the QCD partition function
is violated \cite{lgt1,lgt4}. Consequently the Monte-Carlo method
is not valid and the numerical solution of  the QCD thermodynamics
is not accessible anymore.{\footnote{ The complex structure of the
fermionic contribution to the partition function is very
transparent in the model discussed in the last section. From
Eq.~(\ref{eq64}) it is clear that the complex structure appears
for $\mu_B\neq 0$ since $L_I\neq 0$ for the $SU(N)$ system with
$N>2$. On the other hand the quark generating function (18) is
indeed real for the complex baryon chemical potential.}
%%%%%%%%%%%%%%%%%%%%%%%%%%%f888888888888888888888888
\begin{figure}[htb]
\vskip 1.0 true cm \vskip -0.8cm
\begin{minipage}[t]{60mm}
{\vskip -6.3cm
%{\hskip -0.4cm \includegraphics[width=24.5pc, height=16.3pc]{reduc123.ps}}}\\
{\hskip 0.4cm \includegraphics[width=15.5pc, height=17pc,angle=180]{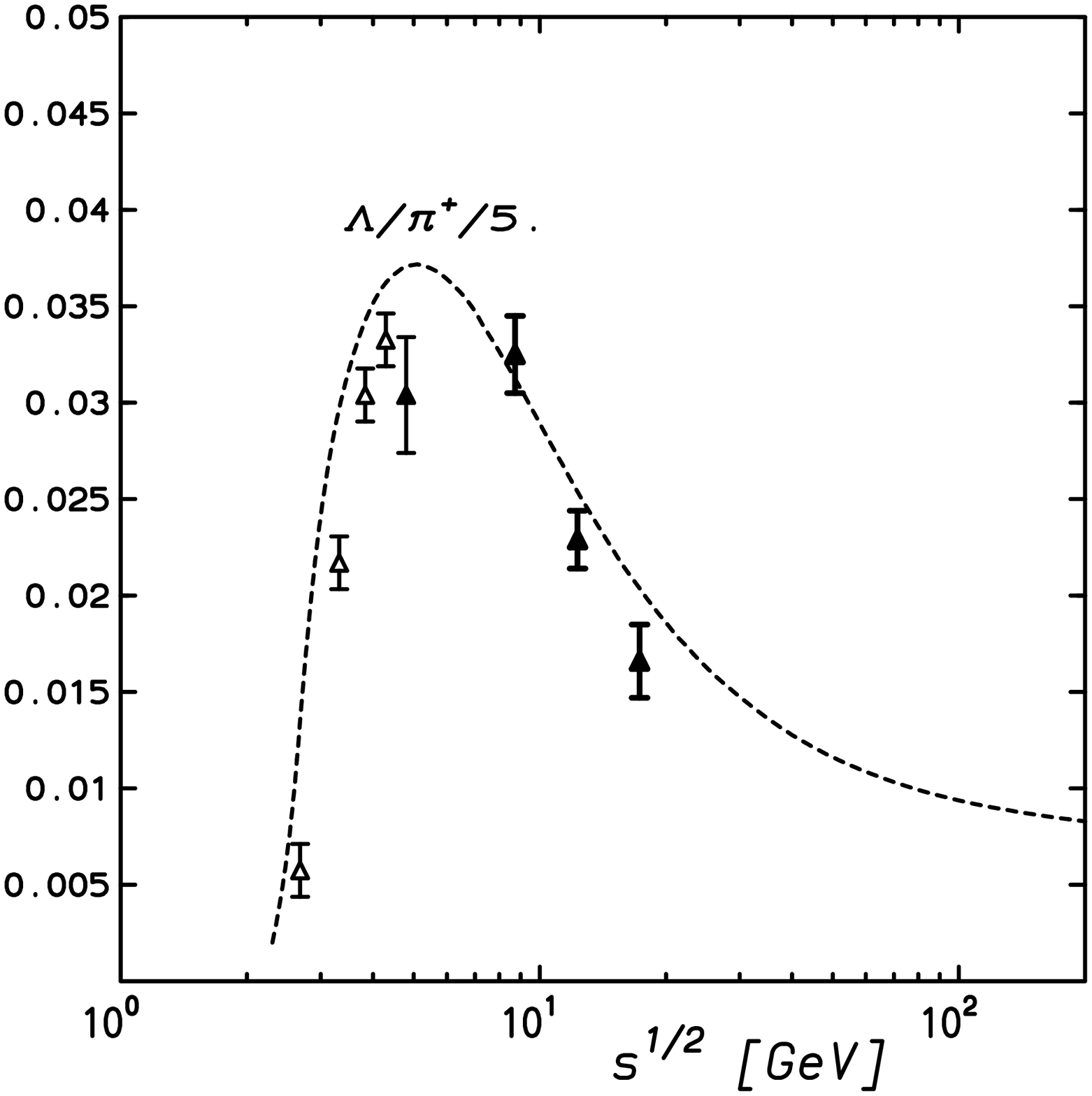}}}\\
\end{minipage}
\hspace{\fill}
\begin{minipage}[t]{65mm}
{%\vskip -6.45cm\hskip -1.6cm
\hskip -0.8cm
\includegraphics[width=15.5pc,height=13.5pc]{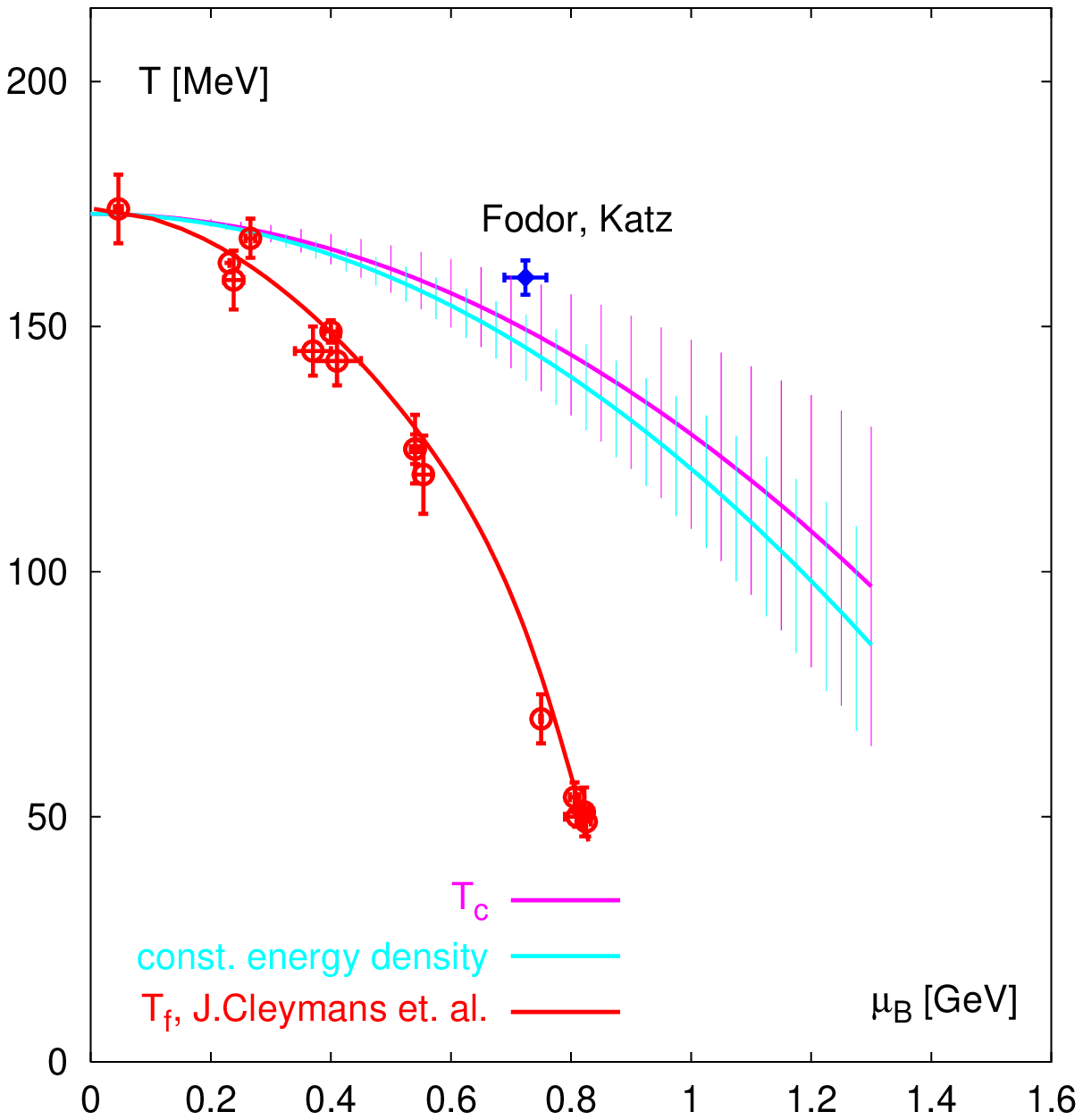}}
\end{minipage}
% \begin{minipage}[t]{197.5mm}
%{\vskip -0.3 true cm \hskip -4.9cm
\vskip -0.7cm\caption{  The left hand figure:
 the  statistical model
results (broken line) on the energy dependence of $\Lambda/\pi$
ratio. The data are from \protect\cite{mischke}. The right hand
figure:
 a comparisons of the chemical freeze-out
parameters  with the phase boundary line. The upper thin line
represents the LGT results obtained in Ref. \protect\cite{r81} and
the lower thin line describes the conditions of constant energy
density that was fixed at $\mu =0$ \protect\cite{r81}. The full ({
Fodor-Katz}) point is the end-point of the crossover transition
from Ref. \protect\cite{r82}. The lower line is the unified
freeze-out curve from \cite{cleymans1}. }
%}
%\end{minipage}
\vskip -0.5cm
\end{figure}
%%%%%%%%%%%%%%%%%%%%%%%%%%%%%%%%%%%%%
One way to partly overcome the above problems in finite density
QCD is  to use a projection method \cite{r64,lw,lgt4}.  Rather
than introducing a non--vanishing chemical potential \cite{lgt2},
that is formulate QCD in the grand canonical ensemble with respect
to U(1)-baryon symmetry, one my go over to a canonical formulation
of the thermodynamics and fix directly the baryon number
\cite{r64,lw,lgt4}. Following the general discussion of Section 1
the projected partition function onto a given sector of baryon
number could be obtained from \cite{r64,lgt4}:

\be Z_B(T,V)=\int_0^{2\pi} {{d\phi}\over {2\pi}} e^{-iB\phi}
Z(\mu_B=i\phi ,T,V) \ee where the  function $Z$ under the  integral
represents the grand canonical partition function
 in Eq.(22) calculated with a complex $\mu_B$, and consequently
  is a real and positive quantity. Nevertheless, due to a   Fourier
 integration  the projected partition function still suffers a
 numerical problem of oscillating functions. However, in the
 quenched limit of QCD and for moderate values of baryon number
  the group  integration  was done analytically
and the first results on QCD thermodynamics at finite baryon
density were  established \cite{lgt4}.

The presence of a finite net baryon number modifies non-trivially
the properties of QCD medium. As an example of Monte--Carlo study
we show in Fig.~2  a heavy quark anti--quark potential obtained
from the Polyakov loop correlations \cite{lgt4}.  For zero baryon
number it shows the usual linearly rising behavior for the
quenched case. For finite $B=6$ the potential remains finite at
large distances due to  screening of the static quark anti--quark
pair  by the already present  net baryon charge in the system.

\subsection{Canonical projection in heavy ion collisions}
Central heavy ion collisions at relativistic incident energies
represent an ideal tool to study nuclear matter at  high
temperatures and densities. Particle production is -- at all
incident energies -- a key quantity to extract information on the
properties of nuclear matter under these extreme conditions. In
this context a  particular role has been attributed  to particles
carrying strangeness that is  related with U(1) invariance of the
strong interactions \cite{r381,peter}. The production  of
secondaries measured in heavy ion collisions was shown in the
literature to be very satisfactory described in the context of
statistical thermal models
\cite{peter,tounsi,braun,r33,braun1,cleymans1}. However, already
the first attempts to describe strange particle production in low
energy central and high energy peripheral heavy ion collisions
have shown that the conservation of strangeness should be
implemented exactly \cite{peter}.

An exact formulation of strangeness conservation is implemented
 through  the projection of the  partition function  onto  states of fixed
 representation of the   U(1)
group \cite{r38}. For the hadron resonance gas the strangeness
neutral partition function is obtained  from \cite{tounsi}:
\begin{equation}\label{ZC}
Z^C_{S=0}(T,V)={1\over {2\pi}}
       \int_{-\pi}^{\pi}
    d\phi~ \exp{\left(\sum_{n=- 3}^3S_ne^{in\phi}\right)}
\end{equation}
where $S_n$   is a thermal  phase space available to all particles
and resonances  that carry strangeness $n$ with $n\in (-3,3) $.
The density $n_s^i$ of particle $i$  carrying strangeness $s$ is
derived form (24)
 as \cite{tounsi}
\begin{equation}
(n_{s}^i)^C= (n_{s}^i)^{GC}F_s(T,V)
\end{equation}
where $(n_{s}^i)^{GC}$ is the grand canonical density and
$F_s(T,V)$ is the  suppression factor that measures a deviation of
$(n_{s}^i)^C$ from its asymptotic,  grand canonical value. For
large $V$ and/or $T$ the factor  $F_s(T,V)\to 1$. The group
projection implies that $F\leq 1$, thus it  suppresses  a particle
densities that carries the U(1) charge. This suppression was found
to increase with decreasing  collision energy, increasing
strangeness content of the particles and decreasing  centrality of
the collisions \cite{tounsi}. These properties are well observed
in  experimental data.  As an illustration for the model
comparison with  experimental data \cite{mischke} we show in
Fig.~(3--left) the  results on the ratio of lambda  to pion
multiplicities \cite{tounsi}.

The statistical model that accounts for an  exact conservation of
U(1) charges was found in the literature to reproduce the basic
features  of particle yields  obtained in heavy ion and
hadron--hadron collisions. The  yields were found to be well
reproduced with  thermal parameters (the temperature and baryon
chemical potential) that follow a universal freeze-out line of
{\it fixed energy per particle} of 1 GeV \cite{cleymans1}. The
freeze-out line is shown in Fig. (3--right) together with the most
recent LGT results on the  position of the critical curve in the
($T$--$\mu_B$) plane \cite{r81,r82}.

\noindent {\bf Acknowledgements:} We wish to thank P.
Braun-Munzinger and L. Turko  for  interesting discussions. K.R
acknowledges the support of the Alexander von Humboldt Foundation.


\vskip 0.2cm
\begin{thebibliography}{000}


%\begin{thebibliography}{99}

\bibitem{r38} K. Redlich and L. Turko,
Z. Phys.  { B97} (1980) 279; L. Turko, Phys. Lett. B104 (1981).

\bibitem{r381}
153; R. Hagedorn and K. Redlich, Z. Phys. { C27} (1985)  541;
Rafelski and M. Danos, Phys. Lett. B97 (1980) 279; L. Turko and J.
Rafelski, Eur. Phys. J. C18 (2001) 587.

\bibitem{r64}
D. Miller and K. Redlich, Phys. Rev. D37 (1988) 3716; Phys. Rev.
D35 (1987) 2524; H. Th. Elze, D. Miller and K. Redlich, Phys. Rev.
D35 (1987) 748.



\bibitem{r64n}
D.H. Rischke, M.I. Gorenstein, A. Sch\"afer, H. St\"ocker and W.
Greiner, Phys. Let. B278 (1992) 19; Z. Phys. C56 (1992) 325.

\bibitem{rgen} A.B. Balantekin, Phys. Rev. E64 (2001) 066105; P.N.
Meisinger, et al., Phys. Rev. D65 (2002) 034009; M.I. Gorenstein,
et al., Phys. Let. B524 (2002) 265; H. Th. Elze, et. al., Phys.
Let. B506 (2001) 123; A.G. Michael, et. al., Phys. Rev. D59 (1999)
034009; M.G. Mustafa, et al., Euro Phys. J. C5 (1988) 711; C.D.
Fosco, Phys. Rev. D57 (1988) 6554; C. Spieles, et. al., Phys. Rev.
C57 (1998) 908; L.D. Mc Lerran and A. Sen, Phys. Rev. D32 (1985)
2794.
\bibitem{peter} P. Braun-Munzinger,  et al.,
% J. Cleymans, H. Oeschler, and
%K. Redlich,
 Nucl. Phys. A697 (2002) 902 and references therein.

\bibitem{tounsi} A. Tounsi, et al.,  J. Phys G28
(2002) 2095; Eur. Phys. J.
 C24 (2002) 35;  J. S. Hamieh, K. Redlich, and A. Tounsi,
 Phys. Lett. B486 (2000) 61
\bibitem{r20}  C.M. Ko, et al.,  Phys. Rev. Lett. 86 (2001)
5438; S. Jeon, et al.,  Nuc. Phys. A697 (2002) 546.
% K. Redlich, V.

\bibitem{quasi}
J. Engels, et al., Z. Phys. C42 (1989) 341; A. Peshier, et al.,
Phys. Rev. C61 (2000) 045203.






\bibitem{lw} A. Roberge and N. Weiss, Nucl. Phys. B275 [FS17]
(1986) 734.

\bibitem{r64p}
A. Gocksch, R.D. Pisarski, Nucl. Phys. B402 (1993) 657; R.
Pisarki, hep-ph/0203271; Nucl. Phys. A702 (2002) 151; A. Dumitru
and R. Pisarski, Phys. Lett. B525 (2002) 95.







\bibitem{lgt} For a recent review see:  I.M. Barbour, S.E.
Morrison, E.G. Klepfish, J.B. Kogut and M.P. Lombardo, Nucl. Phys.
B (proc. Suppl.) 60 (1998) 220.

\bibitem{lgt1}
 F. Karsch, Lect. Notes. Phys. 583
(2002) 209; I. Barbour, et al., Nucl. Phys. Proc. Supp. 60A (1998)
220.

\bibitem{lgt2} P. Hasenfratz and F. Karsch, Phys. Lett. B125
(1983); J. Kogut et al., Nucl. Phys. B225 [FS9] (1983) 93.


\bibitem{lgt3} B. Berg, J. Engels, E. Kehl, B. Waltl and H. Satz,
Z. Phys. C31 (1986) 167.


\bibitem{lgt4}
F. Karsch, Nucl. Phys. Proc. Suppl. 83 (2000) 14; J. Engels et
al., Nucl. Phys. Proc. Suppl. 83 (2000) 366; Nucl. Phys. B558
(1999) 307; O. Kaczmarek et al., Phys. Rev. D62 (2000) 034021.





\bibitem{braun} P. Braun-Munzinger, I. Heppe, and J. Stachel,  Phys.
Lett. B465 (1999) 15;  P. Braun-Munzinger and J. Stachel, J. Phys.
G28 (2002) 1971; Phys. Lett. B490 (2000)196; Nucl. Phys. A690
(2001) 119c; A. Ker\"{a}nen, and F. Becatini, J. Phys.
 G28 (2002) 2041; Becattini, et al.,  Phys. Rev. C64 (2001) 024901.


\bibitem{r33}
 J. Letessier, and J. Rafelski,
{ Int. J.  Mod. Phys.} E9 (2000) 107.
\bibitem{braun1} P. Braun-Munzinger,
et al.,
% D. Magestro, K. Redlich, and
%J. Stachel,
 Phys. Lett. B518 (2001)  41; D. Magestro, J. Phys. G28
(2002) 1745.

\bibitem{cleymans1} J. Cleymans, et al.,
Phys. Rev. C60 (1999) 0544908; Phys. Rev. Lett. { 81} (1998) 5284;
Phys. Rev. { C59} (1999) 1663; Phys. Lett. { B485} (2001) 27.

\bibitem{mischke} A. Mischke, for NA49 Collaboration, nucl-ex/0209002.



\bibitem{r81} S. Ejiri, et al., hep-lat/0209012; C. Schmidt, et
al., hep-lat/0209009.

\bibitem{r82} Z. Fodor and S.D. Katz, Phys. Lett. B534 (2002) 87.




\end{thebibliography}
\end{document}